\documentclass[12pt]{article}
\usepackage{epsfig}
\usepackage{graphicx}
\usepackage{amssymb}
\usepackage{amsmath}
\usepackage{amsfonts}
\usepackage{mathrsfs}
\usepackage[dvips]{color}

\newcommand{\R}{\mathbb{R}}
\newcommand{\C}{\mathbb{C}}

\newcommand{\fa}{\mathfrak{a}}
\newcommand{\fb}{\mathfrak{b}}
\newcommand{\fc}{\mathfrak{c}}
\newcommand{\fd}{\mathfrak{d}}

\newcommand{\fz}{\mathfrak{z}}

\newcommand{\cP}{\mathcal{P}}

\newcommand{\cT}{\mathcal{T}}

\newcommand{\be}{\begin{equation}}
\newcommand{\ee}{\end{equation}}
\newcommand{\bea}{\begin{eqnarray}}
\newcommand{\eea}{\end{eqnarray}}

\newcommand{\ed}{\end{document}}

\newcommand{\bi}{\begin{itemize}}
\newcommand{\ei}{\end{itemize}}

\newcommand{\bce}{\begin{center}}
\newcommand{\ece}{\end{center}}

\newcommand{\RE}{{\rm Re}}
\newcommand{\IM}{{\rm Im}}

\begin{document}

\title{Spectral Singularities of a General Point Interaction}

\author{Ali~Mostafazadeh\thanks{E-mail address:
amostafazadeh@ku.edu.tr, Phone: +90 212 338 1462, Fax: +90 212 338
1559}
\\
Department of Mathematics, Ko\c{c} University, \\ 34450 Sar{\i}yer,
Istanbul, Turkey}

\date{ }
\maketitle

\begin{abstract}
We study the problem of locating spectral singularities of a general
complex point interaction with a support at a single point. We also
determine the bound states, examine the special cases where the
point interaction is $\cP$-, $\cT$-, and $\cP\cT$-symmetric, and
explore the issue of the coalescence of spectral singularities and
bound states. \vspace{2mm}

\vspace{2mm}

\end{abstract}

\section{Introduction}

In elementary courses on quantum mechanics we learn that the
condition that observables are Hermitian operators ensures the
reality of their spectrum. This is an indispensable requirement for
the applicability of the quantum measurement theory, for the points
of the spectrum correspond to readings of a measuring device.
Surprisingly the obvious fact that the reality of the spectrum of an
operator does not necessarily mean that it is Hermitian did not play
much of a role in our understanding of quantum mechanics until the
recent discovery of a large class of  non-Hermitian
$\cP\cT$-symmetric Schr\"odinger operators with a real spectrum
\cite{bender-1998}. This led to a great deal of excitement among
some theoretical physicists. At first it looked as if we could use
this kind of non-Hermitian operators to obtain a quantum theory that
differed (if not generalized or even replaced) the standard quantum
mechanics \cite{bbj-prl-bbj-ajp,bender-review}. But soon it became
clear that the former theory is an equivalent representation of the
latter \cite{review, review-short}. The key was to note that not
only the spectrum but the expectation values, that represented all
the physical quantities, must be real. It turns out that the reality
of expectation values is a stronger condition than the reality of
the spectrum. In fact, it implies that the operator must be
Hermitian (self-adjoint) with respect to the inner product used to
compute the expectation values \cite{review}. This in turn brings up
the possibility of the use of non-standard inner products in quantum
mechanics \cite{p2-p3,bbj-prl-bbj-ajp,jpa-2004,
bender-review,review}. In short, we can use non-Hermitian operators
as observables of a (unitary) quantum system provided that we
Hermitize them by defining the physical Hilbert space appropriately.
In trying to implement this procedure to a delta-function potential
with a complex coupling constant \cite{jpa-2006}, it was noticed
that the standard computational techniques \cite{review} for
determining the inner product (or the corresponding metric operator)
of the physical Hilbert space failed when the coupling constant was
purely imaginary. This marked the occurrence of an intriguing
mathematical phenomenon known as a spectral singularity \cite{math}.
In Ref.~\cite{jpa-2009} we explored this phenomenon for a point
interaction consisting of two delta-function potentials with complex
coupling constants. In Refs.~\cite{prl-2009,pra-2009-pra-2011} we
provided the physical meaning of spectral singularities and outlined
their possible realizations and applications in optics. This
followed by further study of the physical implications of spectral
singularities \cite{others}. The purpose of the present paper is to
examine spectral singularities of a general point interaction with
support at a single point.

First we give the definition of the point interaction in question.

Let $\psi$ be a solution of the time-independent Schr\"odinger
equation
    \be
    -\psi''(x)=k^2\psi(x),~~~~{\rm for}~~~~x\in\R\setminus\{0\},
    \label{sch-eq}
    \ee
$\psi_-$ and $\psi_+$ be respectively the restrictions of $\psi$ to
the sets of negative and positive real numbers, i.e.,
    \be
    \psi_\pm(x):=\psi(x)~~~~{\rm for}~~~~\pm x>0.
    \label{psi-pm=}
    \ee
We also set $\psi_\pm(0):=\lim_{\epsilon\to 0^\pm}\psi_\pm
(\epsilon)$ and introduce the two-component wave function:
    \be
    \Psi_\pm(x):=\left(\begin{array}{c} \psi_\pm(x)\\
    \psi_\pm'(x)\end{array}\right)~~~~{\rm for}~~~~\pm x\geq 0.
    \label{big-psi}
    \ee
Then we can define the point interaction of interest by imposing the
matching condition:
    \be
    \Psi_+(0)= \mathbf{B} \Psi_-(0),~~~~~\mathbf{B}:=
    \left(\begin{array}{cc}
    \fa & \fb \\
    \fc & \fd\end{array}\right),~~~~\fa,\fb,\fc,\fd\in\C.
    \label{B=}
    \ee
Depending on the choice of the coupling constants $\fa,\fb,\fc,\fd$,
this interaction may display $\cP$, $\cT$, or $\cP\cT$-symmetry. The
effects of $\cP\cT$-symmetry on the spectrum of this class of point
interactions have been studied in \cite{afk}. Here we consider the
problem of locating spectral singularities of these interactions. We
will also study the corresponding bound states (eigenvalues with
square-integrable eigenfunctions) using a different approach than
the one pursued in \cite{afk}.

The problem of finding spectral singularities of the above point
interaction turns out not to be as trivial as that of a complex
delta function potential \cite{review-short} and not as complicated
as that of a pair of complex delta function potentials
\cite{jpa-2009}. It is nevertheless exactly solvable and provides a
useful toy model to examine the nature of spectral singularities and
the effects of symmetries on them. For example it allows for the
examination of the possibility of the coalescence of two spectral
singularities.

\section{Spectral Singularities and Bound States}

As discussed in \cite{jpa-2009}, both the spectral singularities and
bound states correspond to zeros of the $M_{22}$ entry of the
transfer matrix $\mathbf{M}$ of the system. For the point
interaction given by~(\ref{B=}),
    \be
    \psi_\pm(x)=A_\pm e^{ikx}+B_\pm e^{-ikx},
    \label{psi-pm}
    \ee
and $\mathbf{M}$ is the $2\times 2$ matrix satisfying
    \be
    \left(\!\!\begin{array}{c} A_+\\
    B_+\end{array}\!\!\right)=\mathbf{M}\left(\!\!\begin{array}{c} A_-\\
    B_-\end{array}\!\!\right).
    \label{M}
    \ee
Combining (\ref{big-psi}) -- (\ref{M}), we find
    \be
    \mathbf{M}=\mathbf{N}^{-1}\mathbf{B\,N}=\frac{-i}{2 k}\left(\begin{array}{cc}
    -\fb k^2+i(\fa+\fd)k+\fc & \fb k^2+i(\fa-\fd)k+\fc \\
    -\fb k^2+i(\fa-\fd)k-\fc & \fb k^2+i(\fa+\fd)k-\fc
    \end{array}\right),
    \label{M=}
    \ee
where
    \[\mathbf{N}:=\left(\begin{array}{cc}
    1 & 1\\ ik & -ik \end{array}\right).\]
According to (\ref{M=}), spectral singularities correspond to real
$k$ values for which
    \be
    \fb k^2+i(\fa+\fd)k-\fc=0,
    \label{zeros}
    \ee
and bound states are given by the complex solutions of this equation
that have a positive imaginary part, \cite{jpa-2009}.

Recall that the transfer matrix for a piecewise continuous
scattering potentials has unit determinant \cite{jpa-2009}. For the
point interaction we consider,
    \be
    \det \mathbf{M}=\det \mathbf{B}=\fa\fd-\fb\fc.
    \ee
Therefore, for the point interactions that are obtained as
``limits'' of a sequence of piecewise continuous functions,
$\fa\fd-\fb\fc=1$. We will call the point interactions violating
this condition \emph{anomalous point interactions}.

In order to examine the solutions of (\ref{zeros}) we consider the
following cases.
    \begin{itemize}
    \item[]\underline{Case I) $\fb\neq 0$:} In this case (\ref{zeros}) gives
        \be
        k=-i(\mu\pm\sqrt{\mu^2-\nu}),
        \label{I}
        \ee
    where
        \be
        \mu:=\frac{\fa+\fd}{2\fb},~~~~~~~~~
        \nu:=\frac{\fc}{\fb}.
        \ee
    Therefore a spectral singularity appears whenever
        \be
        \RE(\mu\pm\sqrt{\mu^2-\nu})=0,
        \label{SS-1}
        \ee
    and a bound state, with a possibly complex energy,
    $k^2=-(\mu\pm\sqrt{\mu^2-\nu}\:)^2$, exists if
        \be
        \RE(\mu\pm\sqrt{\mu^2-\nu})<0.
        \label{BS-1}
        \ee
    The following are some notable special cases:
        \begin{itemize}
        \item[] \underline{I.a) $\fd=-\fa$ (i.e., ${\rm tr}~
    \mathbf{B}=0$):} Then $\mu=0$, $k=\pm\sqrt\nu$ and conditions
    (\ref{SS-1}) and (\ref{BS-1}) become
    $\nu\in\R^+$ and $\IM(\pm\sqrt\nu)<0$, respectively.
        \item[] \underline{I.b) $\fc=0$:} Then $\nu=0$,
    $k=-2i\mu$, we have a spectral singularity
    if $\mu$ is imaginary, and a bound state if $\RE(\mu)<0$.
        \item[] \underline{I.c) $\mu^2=\nu$:} Then the left-hand side of
        (\ref{zeros}) has a double root, namely $k=-i\mu$.
        This corresponds to a spectral singularity, if $\RE(\mu)=0$
        and $\nu\in\R^-$. It gives a bound state of energy
        $k^2=-\mu^2=-\nu$, if $\RE(\mu)<0$.
    \end{itemize}

    \item[]\underline{Case II) $\fb=0$:} In this case we consider the following two possibilities.
            \begin{itemize}
            \item[] \underline{II.a) $\fa+\fd={\rm tr}~\mathbf{B}
            \neq 0$:} Then $k=-i\fc/(\fa+\fd)$, and we have a spectral
            singularity if $\RE(\fc/(\fa+\fd))=0$ and a bound states if
            $\RE(\fc/(\fa+\fd))<0$. A concrete example is the
            delta-function potential with a possibly complex coupling
            constant $\fz$ which corresponds to the choice: $\fa=\fd=1$,
            $\fb=0$, and $\fc=\fz$. These imply
            $\fc/(\fa+\fd)=\fz/2$. Therefore, the system has a spectral
            singularity if $\fz$ is purely imaginary and a bound state
            if $\RE(\fz)<0$, as noted in
            \cite{jpa-2006,review-short}.

            \item[]\underline{II.b) $\fa+\fd={\rm tr}~\mathbf{B}=0$:}
            Then the condition of the existence of a spectral singularity
            or a bound state, namely $M_{22}=0$, implies that $\fc=0$.
            In this case $\mathbf{B}=\fa\,\mathbf{\sigma_3}$ and
            $\mathbf{M}=\fa\,\mathbf{\sigma_1}$, where
            \[\mathbf{\sigma_1}=\left(\begin{array}{cc}
            0 & 1\\ 1 & 0\end{array}\right),~~~~~\mathbf{\sigma_3}=
            \left(\begin{array}{cc} 1 & 0\\ 0 & -1\end{array}\right).\]
            In particular, $\mathbf{M}$ is independent of $k$, $M_{22}$
            vanishes identically, and the interaction is anomalous for
            $\fa\neq\pm i$.
            \end{itemize}
    \end{itemize}

\section{Coalescing Spectral Singularities and Bound States}

Consider case I of the preceding section, i.e., $\fb\neq 0$. Suppose
that $\mu_r:=\RE(\mu)\leq 0$ and $\nu=(1+\frac{\epsilon}{4})\mu^2$
where $\epsilon\in[-1,1]$. Then (\ref{I}) takes the form
    \be
    k=-i\left(1\pm\frac{\sqrt{-\epsilon}}{2}\right)\mu,
    \ee
and we find
    \bea
    \RE(k)=\left\{\begin{array}{ccc}
    \left(1\pm\frac{\sqrt{|\epsilon|}}{2}\right)\mu_i & {\rm for} &
    -1\leq\epsilon <0,\\
    \mu_i & {\rm for} &\epsilon =0,\\
    \mu_i\pm\frac{\sqrt{|\epsilon|}~\mu_r}{2}& {\rm for} &
    0<\epsilon \leq1,
    \end{array}\right.
    \label{eq11}\\
    \IM(k)=\left\{\begin{array}{ccc}
    -\left(1\pm\frac{\sqrt{|\epsilon|}}{2}\right)\mu_r & {\rm for} &
    -1\leq\epsilon <0,\\
    -\mu_r & {\rm for} &\epsilon =0,\\
    -\mu_r\pm\frac{\sqrt{|\epsilon|}~\mu_i}{2}& {\rm for} &
    0<\epsilon \leq1,
    \end{array}\right.
    \label{eq12}
    \eea
where $\mu_i:=\IM(\mu)$.

According to (\ref{eq12}), because $\mu_r\leq 0$ the system has
spectral singularities or bound states. Specifically, if $\mu_r<0$,
then $\IM(k)>0$ and for $\epsilon\in[-1,0)$ there are a pair of
bound states that coalesce into a single bound state at $\epsilon=0$
with energy $-\mu^2$. This marks an exceptional point \cite{ex-pt}.
For $\epsilon\in(0,1]$, the system acquires a pair of bound states,
provided that $|\mu_i\sqrt\epsilon|<-2\mu_r$. If
$|\mu_i\sqrt\epsilon|=-2\mu_r$ then there will be a bound state and
a spectral singularity. If $|\mu_i\sqrt\epsilon|>-2\mu_r$, then
there will be a single bound state as well as a solution of the
Schr\"odinger equation that grows exponentially as $x\to\pm\infty$.

\begin{figure}
    \begin{center}
    \includegraphics[scale=.85,clip]{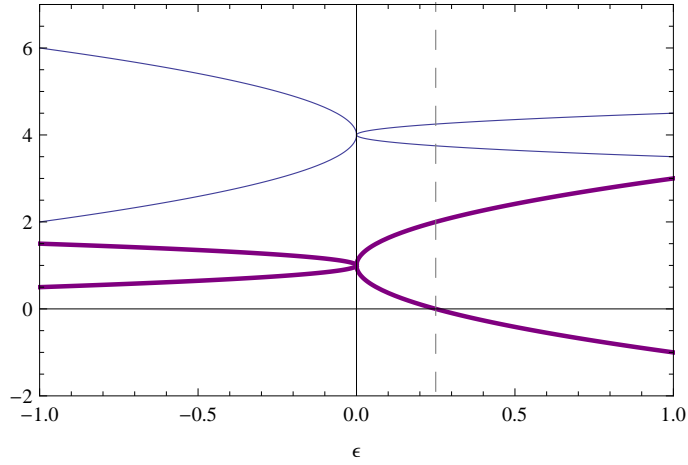}
    \caption{(Color online) Graphs of $\RE(k)$ (the thin blue curve)
    and $\IM(k)$ (the thick purple curve) as a function of
    $\epsilon\in[-1,1]$ for $\mu=-1+4i$,
    $\nu=(1+\frac{\epsilon}{4})\mu^2$. The dashed (grey) line
    is the graph of $\epsilon=\frac{1}{4}$. As $\epsilon$ increases
    starting from $\epsilon=-1$, the initial pair of bound states
    coalesce at $\epsilon=0$ and then split into another pair of bound
    states. One of these only survives until $\epsilon$ reaches the
    critical value $1/4$ at which it turns into a spectral
    singularity and then disappears from the spectrum. \label{fig1}}
    \end{center}
    \end{figure}

Figure~\ref{fig1} shows the plots of $\RE(k)$ and $\IM(k)$ for the
case that $\mu=-1+4i$. As $\epsilon$ changes from $-1$ to $1$, the
two bound states coalesce at $\epsilon=0$ and then split into
another pair of bound states that survive as $\epsilon$ ranges
between 0 and $\frac{1}{4}$. For $\epsilon=\frac{1}{4}$ one of these
turns into a spectral singularity, and for
$\epsilon\in(\frac{1}{4},1)$ it turns into a non-normalizable
solution of the Schr\"odinger equation. The latter corresponds to
the part of the graph of $\IM(k)$ that appears below the
$\epsilon$-axis.

\begin{figure}
    \begin{center}
    \includegraphics[scale=.85,clip]{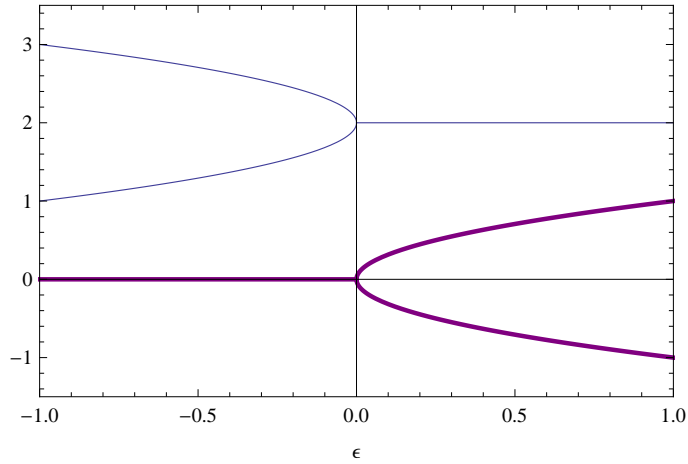}
    \caption{(Color online) Graphs of $\RE(k)$ (the thin blue curve)
    and $\IM(k)$ (the thick purple curve) as a function of
    $\epsilon\in[-1,1]$ for $\mu=2i$ and
    $\nu=-(4+\epsilon)$. As $\epsilon$ increases
    starting from $\epsilon=-1$, the initial pair of spectral
    singularities coalesce at $\epsilon=0$. For $\epsilon>0$ the
    system has a bound state and no spectral singularities.\label{fig2}}
    \end{center}
    \end{figure}

A similar scenario holds for the case that $\mu_r=0$ and $\mu_i\neq
0$. Then for $\epsilon\in[-1,0)$, the system has a pair of spectral
singularities that coalesce at $\epsilon=0$. For $\epsilon\in(0,1]$
the resulting (second order) spectral singularity turns into a bound
state with $k=(1+i\frac{\sqrt\epsilon~}{2})\mu_i$. There also
appears a solution of the Schr\"odinger equation that grows
exponentially as $x\to\pm\infty$. Figure~\ref{fig2} shows this
behavior for $\mu=2i$.

At $\epsilon=0$ both the coalescing bound states and spectral
singularities correspond to a repeated root of $M_{22}$,
equivalently a second order pole of (the singular eigenvalue of) the
$S$-matrix \cite{prl-2009}.

\section{$\cP$, $\cT$, and $\cP\cT$-Symmetries}

In this section we examine the consequences of imposing $\cP$,
$\cT$, and $\cP\cT$-symmetries on the point interaction~(\ref{B=})
and their spectral singularities and bound states.

\subsection{$\cP$-Symmetry}

Let $\cP$ be the parity (reflection) operator acting in the space of
all differentiable complex-valued functions $\psi:\R\to\C$. Then for
all $x\in\R$ we have $(\cP\psi)(x):=\psi(-x)$ and
$(\cP\psi')(x)=-\psi'(-x)$. Therefore in terms of the two-component
wave functions $\Psi:=\left(\begin{array}{c}\psi\\
\psi'\end{array}\right)$, we have
    \be
    (\cP\Psi)(x)=\mathbf{\sigma_3}\Psi(-x).
    \label{P}
    \ee

We say that the point interaction~(\ref{B=}) is $\cP$-invariant (has
$\cP$-symmetry), if
    \be
    (\cP\Psi_+)(0)=\mathbf{B}\,(\cP\Psi_-)(0).
    \ee
We can use this relation and (\ref{P}) to obtain the following
simple expression for $\cP$-invariance of the point
interaction~(\ref{B=}).
    \be
    \mathbf{B\,\sigma_3 B}=\mathbf{\sigma_3}.
    \label{P-sym}
    \ee
In terms of the entries of $\mathbf{B}$, this is equivalent to
    \be
    \fa^2-\fb\fc=1,~~~~\fd=\pm\fa,~~~~
    \fb(\fa-\fd)=\fc(\fa-\fd)=0.
    \label{eq1}
    \ee

Next, consider the problem of spectral singularities and bound
states for $\cP$-invariant point interactions.
     \begin{itemize}
    \item[]\underline{Case I) $\fb\neq 0$:} Then $\fd=\fa$, $\fc=(\fa^2-1)/\fb$, $\mu=\fa/\fb$, $\nu=(\fa^2-1)/\fb^2$, $\mu^2-\nu=1/\fb^2$, and $k=-i(\fa\pm1)/\fb$. Therefore, a spectral singularity exists if $\RE((\fa\pm1)/\fb)=0$, and a bound states arises if $\RE((\fa\pm1)/\fb)<0$.
    \item[]\underline{Case II.a) $\fb=0$ and $\fa+\fd \neq 0$:} Then $\fa=\fd=\pm1$, $k=\mp i\fc/2$, a spectral singularity appears if $\RE(\fc)=0$, and a bound state exists if $\RE(\mp\fc)>0$.
     \item[]\underline{Case II.b) $\fb=0$ and $\fa+\fd = 0$:} Then $\fd=-\fa=\mp 1$, the interaction is anomalous, and spectral singularities and bound states exist for $k\in\R^+$ and $\IM(k)>0$, respectively.
        \end{itemize}

\subsection{$\cT$-Symmetry}

We identify the time-reversal operator $\cT$ as the operator that
acts on complex-valued functions $\psi:\R\to\C$ according to
$(\cT\psi)(x):=\psi(x)^*$. The point interaction (\ref{B=}) is
time-reversal invariant (has $\cT$-symmetry), if
    \be
    (\cT\Psi_+)(0)=\mathbf{B}\,(\cT\Psi_-)(0),
    \ee
where the action of $\cT$ on a two-component wave function $\Psi$ is
defined componentwise. It is easy to see that this relation is
equivalent to the requirement that $\mathbf{B}$ is a real matrix,
i.e., $\fa,\fb,\fc$, and $\fd$ must be real. In this case, we can
summarize the conditions for the existence of spectral singularities
and bound states as follows.
     \begin{itemize}
    \item[]\underline{Case I) $\fb\neq 0$:} Then a spectral singularity
    exists provided that $\mu=0$ and $\nu\in\R^+$.  In terms of the
    entries of $\mathbf{B}$, these relations take the form
        $\fd=-\fa$ and $\fc/\fb\in\R^+$, respectively. The $k$ value
        associated with this spectral singularity is
        $k=\sqrt\nu=\sqrt{\fc/\fb}$. Similarly a bound state with
        $k=\sqrt{\nu-\mu^2}-i\mu$ exists, if $\mu<0$ and $\nu-\mu^2>0$.
        The latter conditions can also be expressed as $(\fa+\fd)/\fb<0$
        and $4\fb\fc-(\fa+\fd)^2>0$, respectively.

    \item[]\underline{Case II.a) $\fb=0$ and $\fa+\fd \neq 0$:} Then no
    spectral singularities exists, and a bound state with
    $k=-i\fc/(\fa+\fd)$ is present provided that $\fc/(\fa+\fd)<0$.
     \item[]\underline{Case II.b) $\fb=0$ and $\fa+\fd = 0$:}
     Then $M_{22}=0$     implies $\fc=0$. As a result, the
     interaction is anomalous, and spectral singularities
     and bound states exist for $k\in\R^+$ and $\IM(k)>0$, respectively.
        \end{itemize}

\subsection{$\cP\cT$-Symmetry}

We say that the point interaction (\ref{B=}) is $\cP\cT$-invariant
or $\cP\cT$-symmetric if
    \be
    (\cP\cT\Psi_+)(0)=\mathbf{B}\,(\cP\cT\Psi_-)(0),
    \ee
This is equivalent to
    \be
    \mathbf{B}^*\mathbf{\sigma_3}\mathbf{B}=\mathbf{\sigma_3}.
    \ee
Expressing this relation in terms of the entries of $\mathbf{B}$ and
solving the resulting equations yield
    \be
    \fa=\sqrt{1+\epsilon_1 bc}\;e^{i\alpha},~~~~
    \fb=\epsilon_1\epsilon_2 b\;e^{i(\alpha+\delta)/2},~~~~
    \fc=\epsilon_2 c\;e^{i(\alpha+\delta)/2},~~~~
    \fd=\sqrt{1+\epsilon_1 bc}\;e^{i\delta},
     \label{PT-sym}
    \ee
where $\epsilon_2=\pm1$, $b,c\in [0,\infty)$,
$\alpha,\delta\in[0,2\pi)$, and
    \be
    \epsilon_1=\left\{\begin{array}{ccc}
    +1&{\rm if}& bc\geq 1,\\
    \pm1&{\rm if}& bc< 1.\end{array}\right.
    \ee
Therefore, the $\cP\cT$-symmetric point interactions are determined
by
    \be
    \mathbf{B}=e^{i(\alpha+\delta)/2}\left(\begin{array}{cc}
        \sqrt{1+\epsilon_1 bc}\;e^{i(\alpha-\delta)/2}&\epsilon_1\epsilon_2 b\\
        \epsilon_2 c &\sqrt{1+\epsilon_1 bc}
        \;e^{-i(\alpha-\delta)/2}\end{array}\right).
        \label{B-PT}
        \ee
In particular, $\det \mathbf{B}=e^{i(\alpha+\delta)}$, and the
interaction is anomalous unless $\alpha+\delta$ is an integer
multiple of $2\pi$.

The conditions for the existence of spectral singularities and bound
states are as follows.
     \begin{itemize}
     \item[]\underline{Case I) $\fb\neq 0$:} A straightforward
     consequence of (\ref{PT-sym}) is that both the parameters $\mu$
     and $\nu$ take real values. More specifically, we have
            \be
            \mu=\frac{\sqrt{1+\epsilon_1 bc}\;
            \cos(\frac{\alpha-\delta}{2})}{\epsilon_1\epsilon_2 b},~~~~~~~
            \nu=\frac{\epsilon_1 c}{b},~~~~~~~
            \mu^2-\nu=\frac{(1+\epsilon_1 bc)
            \cos^2(\frac{\alpha-\delta}{2})-c^2}{b^2}.
            \label{condi}
            \ee
    Now, if we impose the condition of the presence of spectral
    singularities (\ref{SS-1}) we find $\mu=0$ and $\nu>0$. In view
    of (\ref{condi}), these imply that $\epsilon_1=+1$ and
    $\delta=\alpha+(2\ell+1)\pi$, where $\ell$ is an arbitrary
    integer. Therefore, the $\cP\cT$-symmetric point interactions
    that have a spectral singularity (with $k=\sqrt\nu$) are given
    by
        \be
        \mathbf{B}=e^{i\alpha}\left(\begin{array}{cc}
        \sqrt{1+bc}&i\epsilon b\\
        i\epsilon c &-\sqrt{1+bc}\end{array}\right),
        \label{SS-B-PT}
        \ee
    where $\epsilon=(-1)^\ell\epsilon_2=\pm1$. Similarly, we can
    check that the system has a bound state provided that $\mu<0$.
    This is equivalent to
    $\epsilon_1\epsilon_2\cos[(\alpha-\delta)/2]<0$. In this case
    there are three possibilities:
        \begin{itemize}
        \item[]1) $\mu^2-\nu>0$: Then
    there is a pair of bound states with real and negative energies
    $k^2=-(2\mu^2-\nu\pm\mu\sqrt{\mu^2-\nu})$.
        \item[]2) $\mu^2-\nu=0$: This
    corresponds to an exceptional point \cite{ex-pt}, where there is
    single bound state with a real and negative energy $k^2=-\mu^2$.
        \item[]3) $\mu^2-\nu<0$: Then there is a pair of bound states with
    complex-conjugate energies $k^2=\nu-2\mu^2\pm
    i\mu\sqrt{\nu-\mu^2}$.
        \end{itemize}

    \item[]\underline{Case II.a) $\fb=0$ and $\fa+\fd\neq0$:} In
    this case there are no spectral singularities, but a bound state exists provided that $\epsilon_2\cos[(\alpha-\delta)/2]<0$. It has a real and
    negative energy given by
    $k^2=-\frac{1}{4}c^2\sec^2[(\alpha-\delta)/2]$.

    \item[]\underline{Case II.b) $\fb=0$ and $\fa+\fd=0$:} Then the
    condition $M_{22}=0$ implies $c=0$,
        \[B=\left(\begin{array}{cc} e^{i\alpha} & 0\\
        0 & -e^{i\alpha}\end{array}\right),\]
    the interaction is anomalous unless $\alpha=\frac{\pi}{2}$ or
    $\frac{3\pi}{2}$, a spectral singularity arises
    for all $k\in\R$, and there is a bound state for all $k\in\C$ with
    $\IM(k)>0$.

    \end{itemize}

    \section{Concluding Remarks} In this article we present a
    complete solution for the problem of spectral singularities of a
    general point interaction with a support at a single point. We
    also examine the consequences of the presence of $\cP$-, $\cT$-,
    and $\cP\cT$-symmetries. Unlike the case of complex delta-function
    potential and double-delta function potential, for a generic point
    interaction that we consider, the function whose zeros give the
    spectral singularities and bound states is a quadratic polynomial
    in $k$. This in turn implies the possibility of having a spectral
    singularity or a bound state that is related to a second order zero of this polynomial.  For the case of a bound state this corresponds to a degeneracy
    or exceptional point. The latter leads to a well-known type of geometric phases \cite{ex-pt}. The analogy with coalescing spectral singularities calls for a thorough examination of the geometric phase problem for systems supporting second and higher order spectral singularities.

\vspace{.5cm} \noindent {\em Acknowledgments:} This work has been
supported by  the Scientific and Technological Research Council of
Turkey (T\"UB\.{I}TAK) in the framework of the project no: 110T611
and the Turkish Academy of Sciences (T\"UBA).


\begin{thebibliography}{99}

\bibitem{bender-1998} C.~M.~Bender and S.~Boettcher,
Phys.\ Rev.\ Lett.\ {\bf 80}, 5243 (1998).

\bibitem{bbj-prl-bbj-ajp} C.~M.~Bender, D.~C.~Brody and H.~F.~Jones, Phys.\ Rev.\ Lett.\ {\bf 89}, 270401 (2002), Erratum Phys.\ Rev.\ Lett.\ {\bf 92}, 119902 (2004);\\
    C.~M.~Bender, D.~C.~Brody and H.~F.~Jones, Am.~J.~Phys.\ {\bf 71}, 1095 (2003).

\bibitem{bender-review} C.~M.~Bender, Rep.\ Prog.\ Phys.~{\bf 70},
947 (2007).

\bibitem{review} A.~Mostafazadeh, Int.\ J.~Geom.\ Meth.\ Mod.\ Phys.~\textbf{7}, 1191 (2010); arXiv:0810.5643.

\bibitem{review-short} A.~Mostafazadeh, Phys.\ Scr.\ 82, 038110 (2010).

\bibitem{jpa-2003-cjp-2003} A.~Mostafazadeh, J.~Phys.~A {\bf 36}, 7081
(2003);\\ A.~Mostafazadeh, Czech J.~Phys.\ {\bf 54}, 1125 (2004).

\bibitem{p2-p3} A.~Mostafazadeh, J.\ Math.\ Phys.\ {\bf 43}, 205
(2002); ibid 2814 (2002); ibid 3944 (2002).

\bibitem{jpa-2004} A.~Mostafazadeh and A.~Batal,
J.~Phys.~A {\bf 37}, 11645 (2004).

\bibitem{jpa-2006} A.~Mostafazadeh, J.~Phys.~A {\bf 39}, 13495 (2006).

\bibitem{math} M.~A.~Naimark, Trudy Moscov.\ Mat.\ Obsc.\ \textbf{3}, 181 (1954) in Russian, English translation: Amer.\ Math.\ Soc.\ Transl.\
(2), \textbf{16}, 103 (1960);\\
 R.~R.~D.~Kemp, Canadian J. Math.
\textbf{10}, 447 (1958);\\ J.~Schwartz, Comm.\ Pure Appl.\ Math.
\textbf{13}, 609 (1960);\\ G.~Sh.~Guseinov, Pramana.\ J.~Phys.\
\textbf{73}, 587 (2009).

\bibitem{jpa-2009} A.~Mostafazadeh and H.~Mehri-Dehnavi,
J.~Phys.~A \textbf{42}, 125303 (2009).

\bibitem{prl-2009} A.~Mostafazadeh, Phys.\ Rev.\ Lett.~\textbf{102}, 220402
(2009).

\bibitem{pra-2009-pra-2011}
    A.~Mostafazadeh, Phys.\ Rev.\ A \textbf{80}, 032711
(2009); ibid  \textbf{83}, 045801 (2011).

\bibitem{others} S.~Longhi, Phys.\ Rev.\ B  \textbf{80}, 165125 (2009) and Phys.\ Rev.\ A \textbf{81}, 022102 (2010);\\  B.~F.~Samsonov, preprint arXiv:1007.4421.

\bibitem{afk} S.~Albeverio, S.-M.\ Fei, and P.~Kurasov, Lett.\ Math.\ Phys.\ {\bf 59}, 227 (2002).

\bibitem{ex-pt}
W.~D.~Heiss, M.~M\"uller, and I.~Rotter, Phys.\ Rev.~E
{\bf 58}, 2894 (1998);\\ W.~D.~Heiss, J.~Phys.\ A {\bf 37}, 2455 (2004);\\
T.~Stehmann, W.~D.~Heiss, and F.~G.~Scholtz, J.~Phys.\ A {\bf 37},
7813 (2004);\\ C.~Dembowski, B.~Dietz, H.-D.~Gr\"{a}f, H.~L.~Harney,
A.~Heine, W.~D.~Heiss, and A.~Richter, Phys.\ Rev.~E {\bf 69},
056216 (2004);\\ A. A. Mailybaev, O. N. Kirillov, and A. P.
Seyranian, Phys.\ Rev.\ A., {\bf 72}, 014104 (2005);\\
H. Mehri-Dehnavi and A. Mostafazadeh, J.~Math.\ Phys.\ \textbf{49},
082105 (2008).


\end{thebibliography}
\end{document}